\documentclass[journal, twoside]{IEEEtran}
\usepackage{indentfirst}
\usepackage{graphicx}
\usepackage{amsmath}
\usepackage{amssymb}
\usepackage{amsfonts}
\usepackage{mathrsfs}
\usepackage{leftidx}
\usepackage{color}
\usepackage{amsmath}
\usepackage{arydshln}
\usepackage{amsthm}
\usepackage{ragged2e}
\usepackage{cite}
\usepackage{enumerate}
\usepackage{longtable}
\usepackage{float}
\usepackage{stfloats}
\usepackage{hyperref}
\usepackage{algpseudocode}
\usepackage{algorithm}
\usepackage[caption=false,font=footnotesize]{subfig}
\usepackage{tabularx}
\usepackage{makecell}
\usepackage{url}

\usepackage{multirow} 
\usepackage{booktabs}
\theoremstyle{plain}

\usepackage{caption}

\newcolumntype{P}[1]{>{\raggedright\arraybackslash\footnotesize}m{#1}}
\newcolumntype{A}[1]{>{\centering\arraybackslash\footnotesize}m{#1}}

\usepackage[table,usenames,dvipsnames]{xcolor}
\definecolor{aa}{RGB}{175,238,238}
\definecolor{bb}{RGB}{255,255,255}

\usepackage{bm}
\usepackage{makecell}

\begin{document}

\title{Secure Semantic Communication With Homomorphic Encryption}

\author{Rui Meng,~\IEEEmembership{Member,~IEEE,} Dayu Fan, Haixiao Gao, Yifan Yuan, Bizhu Wang, 

Xiaodong Xu,~\IEEEmembership{Senior Member,~IEEE,} 
Mengying Sun,~\IEEEmembership{Member,~IEEE,} Chen Dong, 

Xiaofeng Tao,~\IEEEmembership{Senior Member,~IEEE,} 
Ping Zhang,~\IEEEmembership{Fellow,~IEEE,} and Dusit Niyato,~\IEEEmembership{Fellow,~IEEE}

\thanks{
\textit{Corresponding author: Xiaodong Xu.}

Rui Meng, Dayu Fan, Haixiao Gao, Yifan Yuan, Bizhu Wang, Xiaodong Xu, Mengying Sun, Chen Dong, and Ping Zhang are with the State Key Laboratory of Networking and Switching Technology, Beijing University of Posts and Telecommunications, Beijing 100876, China (e-mail: buptmengrui@bupt.edu.cn; fandayu@bupt.edu.cn; haixiao@bupt.edu.cn; yifanyuan@bupt.edu.cn; wangbizhu\_7@bupt.edu.cn; xuxiaodong@bupt.edu.cn; smy\_bupt@bupt.edu.cn; dongchen@bupt.edu.cn; pzhang@bupt.edu.cn).

Xiaofeng Tao is with the National Engineering Laboratory for Mobile Network Technologies, Beijing University of Posts and Telecommunications, Beijing 100876, China (email: taoxf@bupt.edu.cn).

Dusit Niyato is with College of Computing and Data Science, Nanyang Technological University, Singapore (email: dniyato@ntu.edu.sg).
}}

\maketitle

\begin{abstract}
In recent years, Semantic Communication (SemCom), which aims to achieve efficient and reliable transmission of meaning between agents, has garnered significant attention from both academia and industry. To ensure the security of communication systems, encryption techniques are employed to safeguard confidentiality and integrity. However, existing encryption schemes encounter obstacles when applied to SemCom. To address this issue, this paper explores the feasibility of applying homomorphic encryption (HE) to SemCom. Initially, we review the encryption algorithms utilized in mobile communication systems and analyze the challenges associated with their application to SemCom. Subsequently, we overview HE techniques and employ scale-invariant feature transform (SIFT) to demonstrate that the extractable semantic information can be preserved in homomorphic encrypted ciphertext. Based on this finding, we further propose the HE-joint source-channel coding (HE-JSCC) scheme, where the traditional JSCC model architecture is modified to support HE operations. Moreover, we present the simulation results for image classification and image generation tasks. Furthermore, we provide potential future research directions for homomorphic encrypted SemCom.


\end{abstract}

\section{Introduction}
As one of the most promising technology for future \textit{intellicise (intelligent and concise)} wireless networks, Semantic Communication (SemCom) initially extracts, compresses, and transmits the selective features of the original signal, and subsequently employs semantic information to facilitate effective communication \cite{zhang2024intellicise}. This innovative ``comprehend before transmitting" paradigm allows both communicating parties to transmit information based on their needs, effectively eliminating redundant elements. Consequently, this approach leads to a substantial decrease in the volume of information transmitted and a notable reduction in energy consumption \cite{meng2025survey}.

As an emerging technology, SemCom faces significant security challenges, such as poisoning or backdoor attacks to knowledge bases \cite{shen2023secure}, semantic eavesdropping attacks \cite{meng2025survey}, and semantic inference attacks \cite{yang2024secure}.
In addressing these security threats, encryption techniques serve to safeguard confidentiality and integrity. Tung et al. \cite{tung2023deep} introduced the first secure SemCom scheme for wireless image transmission, which protects against eavesdropping attacks by providing resilience to chosen-plaintext attacks without assumptions about the eavesdropper's channel conditions. Recently, researchers have integrated additional technologies to protect the privacy of semantic information, such as physical layer semantic encryption \cite{chen2024lightweight}, adversarial encryption training \cite{luo2023encrypted}, and semantic steganography \cite{wang2025image,gao2025semstediffgenerativediffusionmodelbased}.
Nonetheless, research on encrypting SemCom remains in its nascent stages. 
When directly applied to SemCom, the state-of-the-art solutions still face challenges, such as relying on upper layer security protocols or causing new security threats.
Besides, the application of some emerging privacy protection technologies, such as homomorphic encryption (HE), has yet to be thoroughly explored. Specifically, the following questions have not been answered:

\textit{Q1: What are the challenges of encrypting SemCom?}

\textit{Q2: Is HE suitable for SemCom?}

\textit{Q3: What new issues does HE enabled SemCom bring?}

Motivated by solving these challenges, we evaluate the feasibility of applying HE to SemCom. The main contributions are summarized as follows.
\begin{itemize}
\item We revisit classical encryption techniques for traditional mobile communication systems, and analyze the limitations of applying existing encryption schemes to SemCom (For Q1).
\item To overcome these challenges, we believe that the homomorphic operation nature of HE could be advantageous for semantic processing in joint source-channel coding (JSCC). Firstly, we utilize Scale-Invariant Feature Transform (SIFT) to verify that the homomorphic encrypted ciphertext still retains extractable semantic information. Then, we analyze how to modify the JSCC architecture to support HE operations, and propose the HE-JSCC scheme. Furthermore, we provide the simulation results based on open-source datasets for different communication tasks (For Q2).
\item We envision the future research direction of SemCom with HE empowerment (For Q3).
\end{itemize}

\begin{figure*}[tbp]
\centering
    \subfloat[E2E traditional communication framework.]{\includegraphics[width=88mm]{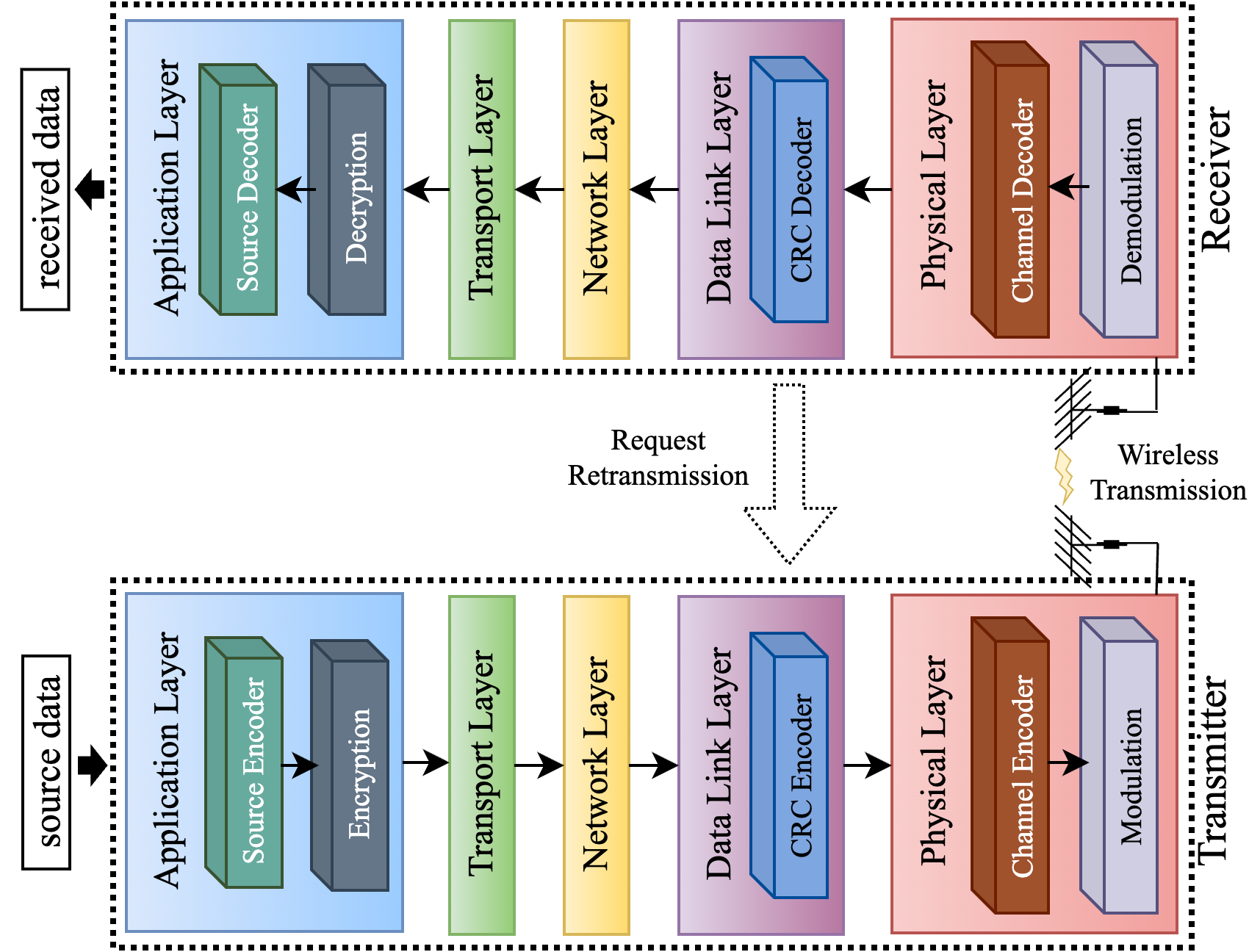}}%
    \hfill
    \subfloat[E2E semantic communication framework.]{\includegraphics[width=88mm]{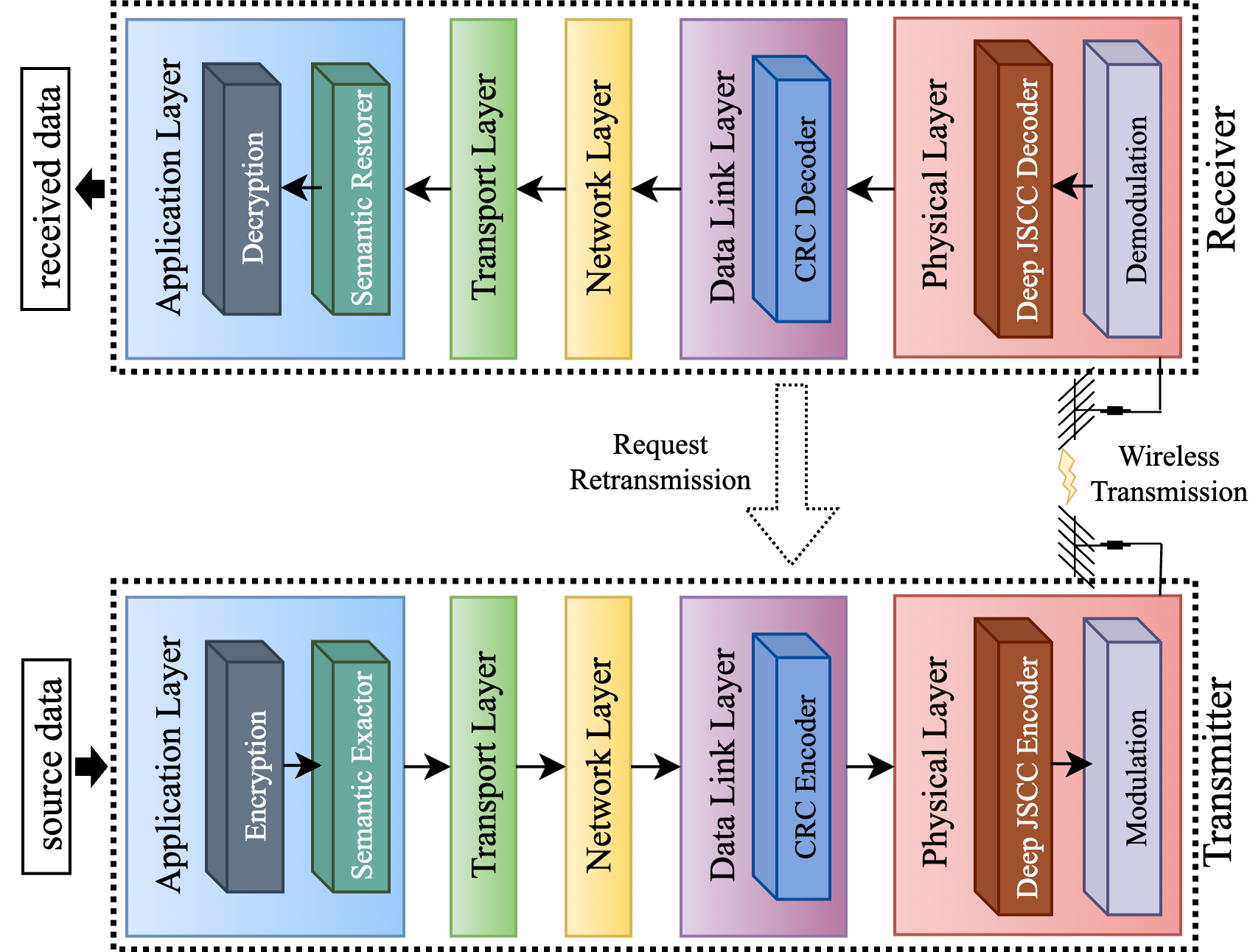}}%
  
    \caption{A comparison between traditional E2E communication framework and semantic communication (SemCom) framework. Both correspond to the TCP/IP five-layer structure, which includes the application layer, transport layer, network layer, data link layer, and physical layer. (a) Traditional communication employs a separate source and channel coding approach, conducting source coding and encryption at the application layer, and channel coding at the physical layer to reduce the bit error rate. (b) SemCom uses joint source-channel coding (JSCC), where information is encrypted and semantically compressed at the application layer, then combined with channel information for coding at the physical layer.}
    \label{fig1}
\end{figure*}

\section{Overview of Semantic Communication (SemCom) and Encryption Techniques}
This section first compares traditional communication and SemCom frameworks, then reviews encryption techniques for traditional communication systems, and finally analyzes the challenges of encrypting SemCom.
\subsection{What is SemCom?}
Figure \ref{fig1} compares the end-to-end (E2E) traditional and SemCom frameworks, with the main differences discussed as follows. 1) Although both frameworks perform encryption and decryption at the application layer, SemCom systems require the extraction and restoration of semantic features after encryption and before decryption, respectively\footnote{It is noted that perform encryption prior to semantic extraction, enabling the entire semantic encoding/decoding and channel robustness optimization to be completed within the ciphertext domain. This approach simultaneously achieves end-to-end confidentiality and reliable transmission. In contrast, the method of ``extracting semantics first and then encrypting" either leads to greater security exposure or scrambles the feature stream into random bits, undermining subsequent robust transmission and inference.}. This is unnecessary in traditional communication systems. 
2) In the traditional communication framework, the channel decoder is only responsible for performing error correction on the received bits at the physical layer, and its goal is to recover the bit sequence transmitted by the source as accurately as possible. In contrast, in the SemCom architecture, JSCC\footnote{It is noted that JSCC is only one way to realize SemCom. The reason why we adopt JSCC is that, our objective is to verify that semantic-level processing in the ciphertext domain is feasible. Current HE schemes are efficient only for addition and multiplication. The JSCC architecture can be systematically modified into an HE-friendly E2E pipeline. For more details, see Section IV.} jointly models semantic compression and channel robustness through deep learning techniques, thereby directly performing E2E transmission and recovery of semantic features at the physical layer. This design overcomes the ``cliff effect" inherent in traditional systems and significantly enhances transmission performance in low signal-to-noise ratio (SNR) environments.

\subsection{What are Encryption Techniques for Traditional Mobile Communication System?}
\subsubsection{Cryptography-based Encryption Techniques}
Encryption algorithms are mathematical methods used to convert (encrypt) data into an unreadable or difficult-to-understand format, thereby protecting the confidentiality of the data. Only legitimate users with the correct key can decrypt the data, restoring it to its original and understandable form. 
Table \ref{tab1} illustrates the evolution of encryption algorithms in mobile communication systems from 1G to 5G. 
Currently, to achieve confidentiality and integrity protection, 5G employs New radio Encryption Algorithms (NEA) and New radio Integrity Algorithms (NIA), both of which are based on the SNOW 3G, AES, and ZUC algorithms that were also utilized in 4G.

\begin{table*}[tbp] 
    \centering
    \caption{The evolution of encryption algorithms of mobile communication systems}
    \label{tab1}
    \renewcommand{\arraystretch}{1.2}  
    \normalsize  
    \begin{tabular}{|>{\centering\arraybackslash}m{0.03\textwidth}|>{\centering\arraybackslash}m{0.09\textwidth}|>{\arraybackslash}m{0.82\textwidth}|}
        \hline
        \textbf{\small{Syst.}} & \textbf{\small{Encryption Algorithm}} & \textbf{\small{Description}} \\
        \hline
        \small{1G} & \small{Not encrypted} & \small{1G primarily employs analog technology. Voices are transmitted through unencrypted radio waves with a transmission rate of approximately 2.4 Kbps.} \\
        \hline
        \small{2G} & \small{A5} & \small{1. Four versions are developed, namely A5/1, A5/2, A5/3, and A5/4.
        
        2. The shift registers are too short, making them susceptible to exhaustive attacks. In the evolution of communication systems, its security is continually being challenged.} \\
        \hline
        \small{3G} & \small{KASUMI, 
        
        SNOW 3G}
        
        & \small{1. The KASUMI algorithm is a block cipher incorporating security measures against differential attacks.
        
        2. The SNOW 3G algorithm is a byte-oriented stream cipher that generates processed byte streams based on input parameters, with a one-to-one correspondence between inputs and outputs.
        
        3. Compared to the KASUMI algorithm, the SNOW 3G algorithm can process data streams more efficiently and offers higher data throughput and better compatibility. The SNOW 3G algorithm has been widely adopted in 3G and subsequent mobile communication systems, while the KASUMI algorithm is primarily used in 3G systems.} \\
        \hline
        \small{4G} & \small{SNOW 3G, 
        
        AES, ZUC}
        
        & \small{1. Compared to the KASUMI algorithm, the AES algorithm undergoes extensive security verification, supports multiple key lengths, and offers higher computational efficiency and stronger compatibility. Consequently, the KASUMI algorithm is superseded by the AES algorithm. 
        
        2. The ZUC algorithm is an encryption algorithm independently developed by China. In the LTE Rel-11 version, the ZUC algorithm becomes the third optional encryption algorithm.} \\
        \hline
        \small{5G} & \small{SNOW 3G, 
        
        AES, ZUC}
        
        & \small{1. The encryption algorithms used for confidentiality protection are known as NEA algorithms, which include NEA1 based on the SNOW 3G algorithm, NEA2 based on the AES algorithm, and NEA3 based on the ZUC algorithm. NEA facilitates secure transmission between the sender and receiver, guaranteeing end-to-end data confidentiality. By utilizing a 128-bit key length, it offers superior security compared to 64-bit encryption algorithms, rendering encrypted data more resilient to cracking attempts. 
        
        2. 5G employs NIA algorithms to safeguard the integrity of the air interface, utilizing the same key length and underlying encryption algorithm as those employed for confidentiality protection. Through the calculation of the message authentication code and expected message authentication code, it swiftly detects tampering or damage to data, thereby ensuring the integrity of transmitted information.} \\
        \hline
    \end{tabular}

\end{table*}




\subsubsection{Physical Layer Security (PLS) Techniques}

PLS techniques are grounded in information theory and are designed to protect against eavesdroppers by utilizing the inherent properties of the wireless transmission medium. It is an important endogenous security supplement to cryptography-based encryption techniques implemented in the upper layer.
PLS technology encompasses physical layer key generation and covert communications. The former relies on channel fading, noise, or unique device attributes. These characteristics are both random and unique, imparting a distinct ``fingerprint" to each wireless link.
The latter, covert communication, aims to ensure that information is securely transmitted without detection by an eavesdropper. Its implementation hinges on various inherent or artificial uncertainties. These PLS methods can also be integrated with other physical-layer technologies, such as Intelligent Reconfigurable Surface (IRS) and beamforming, to further enhance the security.


\subsection{What are the Challenges of Encrypting SemCom?}

We analyze the challenges of encrypting SemCom as follows.
\begin{itemize}
\item \textbf{The Inherent Contradiction between Semantic Processing and Traditional Encryption Mechanisms:} 
Traditional encryption methods typically convert plaintext data into random ciphertext to ensure the confidentiality of information. However, this transformation destroys the semantic correlation in the data, making it difficult to extract semantic information from ciphertext and limiting the ability of SemCom systems to understand and utilize semantic information \cite{li2024toward}. In addition, SemCom requires operations such as data pre-processing, feature representation, feature selection, and parameter updates to extract, transmit, and recover semantic information. However, the ciphertext generated by traditional encryption algorithms can not directly perform complex arithmetic and logical operations other than decryption, which hinders semantic analysis and processing \cite{guo2024survey}. Although PLS technology can enhance the security of communication systems, it can not replace upper layer encryption technology and still can not solve the challenges brought by traditional cryptographic algorithms.
\item \textbf{New Security Threats Brought by DL-based Encryption Technology:} Although researchers have designed DL-based encryption techniques, such as adversarial training \cite{luo2023encrypted} and Generative Artificial Intelligence (GAI) \cite{wang2025image,gao2025semstediffgenerativediffusionmodelbased}, they may introduce new security threats to SemCom. Adversarial training improves the robustness of JSCC models by generating adversarial samples, but attackers may exploit vulnerabilities not covered by the model to design new semantic adversarial samples, making existing methods difficult to deal with complex and ever-changing threats. Besides, adversarial training requires a large amount of data to train JSCC models. If there is a lack of protection during data collection, storage, or transmission, it may be intercepted by attackers and used for reverse engineering JSCC models \cite{meng2025survey}. Although GAI can generate realistic text, images, or audio based on semantic information, it may be used to forge false information and mislead legitimate recipients. Furthermore, the decision-making process of GAI is complex and difficult to explain, especially in safety critical areas such as remote healthcare, where semantic confusion may lead to generating incorrect content.
\end{itemize}


\begin{figure*}
\centering
\includegraphics[width=1\textwidth]{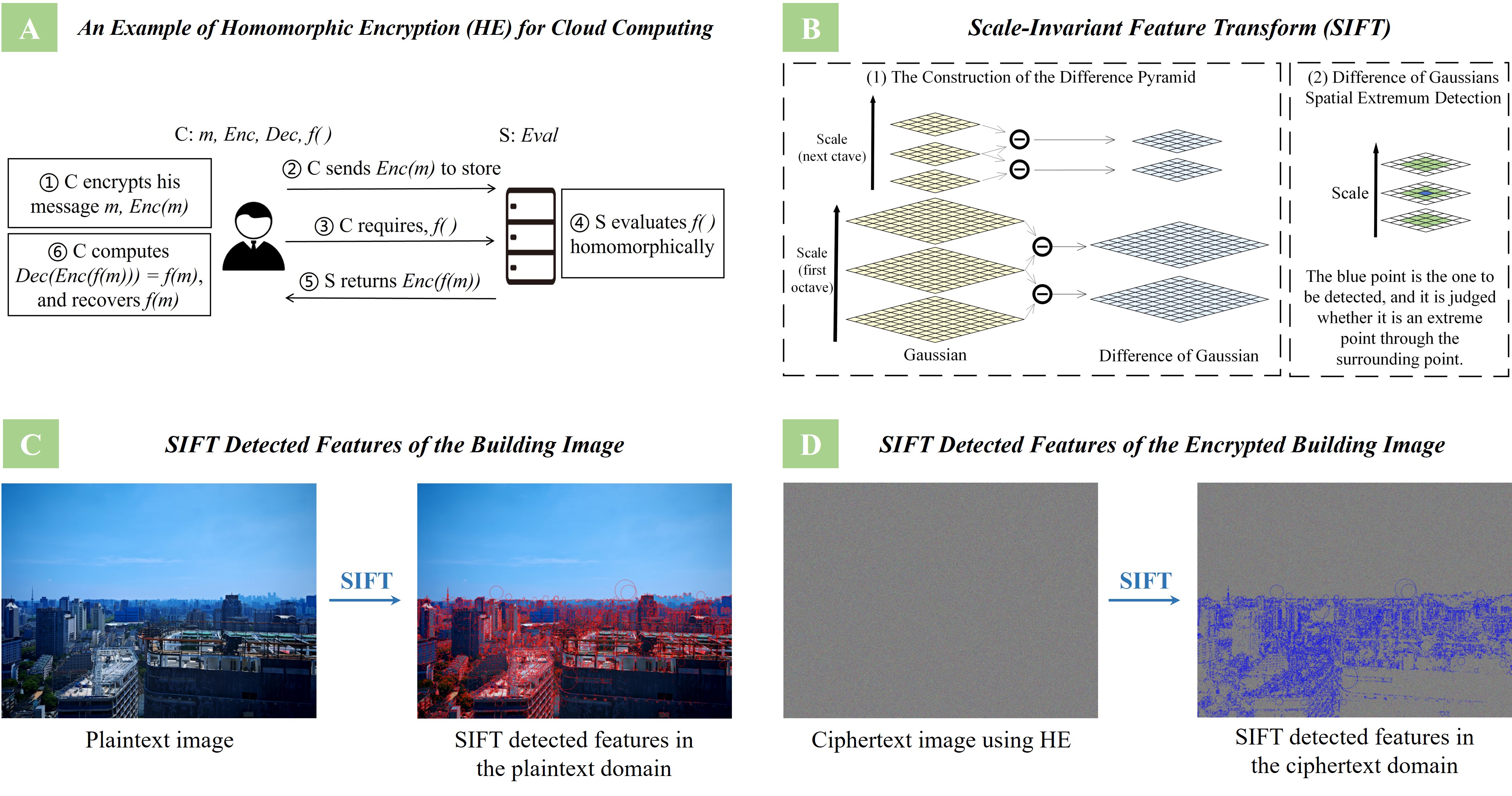}
\caption{Verification that homomorphic encrypted ciphertext still retains extractable semantic information. \textit{Part A} shows an example of HE for cloud computing. \textit{Part B} illustrates the steps of SIFT algorithm (difference-of-Gaussian transforms, scale-space extrema detection, and keypoint localization). \textit{Part C} illustrates the SIFT detected features of a plaintext image. \textit{Part D} illustrates the SIFT detected features of a homomorphic encrypted image.}
\label{fig2}
\end{figure*}

\section{Motivations of Applying Homomorphic Encryption (HE) to Secure SemCom}
This section first overviews HE techniques, then combines SIFT to verify the potential of HE in encrypting SemCom, and further analyzes what SemCom can benefit from HE.

\subsection{Overview of HE}
As illustrated in Figure \ref{fig2} \textit{Part A}, HE is a distinctive encryption mode in cryptography, referring to encryption algorithms that adhere to the principle of homomorphic operations on ciphertexts \cite{falcetta2022privacy}. This encryption method enables computations to be conducted directly on encrypted data without requiring decryption first, ensuring that the decrypted outcome mirrors the result obtained by performing the same operations on plaintext data. HE thus achieves the effect of ``computation without visibility", allowing processors to manipulate ciphertexts and yield encrypted results while preserving the confidentiality of the underlying data.

HE can be categorized into fully HE (FHE) and partially HE (PHE). The former supports a wide range of computations on ciphertexts, whereas the latter restricts these operations to specific types or a limited combination of such operations.
Representative FHE schemes include Yet Another Somewhat HE (YASHE), Brakerski/Fan-Vercauteren (BFV), Brakerski-Gentry-Vaikuntanathan (BGV), Cheon-Kim-Kim-Song (CKKS), and Fast Fully HE over the Torus (TFHE). YASHE, BFV, and BGV are designed for integer arithmetic, supporting only addition and multiplication operations. In contrast, CKKS specializes in real-number computations, enabling addition, multiplication, and division. TFHE, on the other hand, is tailored for binary operations, supporting bitwise manipulations. 

\subsection{What can SemCom Benefit from HE?}

The JSCC module in SemCom is typically implemented with neural networks, which are known for their robust feature extraction capabilities. Neural networks abstract input data through a multi-layer structure, thus extracting high-level semantic features. If these networks adhere to the principles of homomorphic computation, then HE holds significant potential for overcoming the challenges posed by encrypting SemCom.

Here, we utilize Scale-Invariant Feature Transform (SIFT) to demonstrate that the ciphertext resulting from HE retains extractable semantic information. As illustrated in Figure \ref{fig2} \textit{Part B}, we use the SIFT algorithm proposed by \cite{hsu2012image}. It constructs a scale-space to analyze images at various scales and identify keypoints.
These keypoints are extrema points within the scale-space, exhibiting invariance to location, scale, and rotation.
In semantic segmentation, scale invariance is a critical property as it ensures that objects or features of different scales are accurately recognized and processed during segmentation. The scale invariance of SIFT allows it to detect identical feature points in images captured at varying resolutions and locations, enabling SIFT to extract semantic features from images effectively.
Figure \ref{fig2} \textit{Part C} displays the photograph of the building that we captured and the corresponding detected SIFT features. Figure \ref{fig2} \textit{Part D} shows the encrypted version using HE. By comparing Figure \ref{fig2} \textit{Part C} and \textit{Part D}, it is evident that the detected semantic features (corner points, edge points, bright spots in dark areas, and dark spots in bright areas) in both the plaintext and ciphertext domains are almost visually identical. Therefore, the ciphertext resulting from HE retains semantic information. Based on this finding, we conclude that SemCom can benefit from HE for the following aspects:
\begin{itemize}
\item \textbf{Complete Protection:} In SemCom, HE is embedded in the application layer processing flow to form E2E encryption protection, as illustrated in Figure \ref{fig1}. Specifically, after the source data is encrypted and packaged at the application layer of the transmitter, it will traverse the entire communication link in ciphertext form until the decryption operation is performed at the application layer of the receiver. This design mechanism enables the privacy information to always remain in an encrypted form in each layer of the SemCom system, providing comprehensive protection.
\item \textbf{Avoidance of Avalanche Effects:} 
As demonstrated before, semantic information can be reliably extracted from homomorphically encrypted ciphertexts. If the JSCC framework supports homomorphic operations, this intrinsic property ensures that the semantic features of source data remain intact, thereby avoiding avalanche effects of traditional encryption algorithms.
\end{itemize}

\section{Proposed HE-JSCC Scheme for Secure SemCom}
This section first introduces the designed HE-JSCC model, then provides the steps of the HE-JSCC scheme and analyzes its advantages, further presents the simulation results under image classification and image generation tasks.

\begin{figure*}[]
\centering

\subfloat[The designed HE-JSCC model, include the activation function, convolutional layer, pooling layer, and batch normalization.]{\includegraphics[width=140mm]{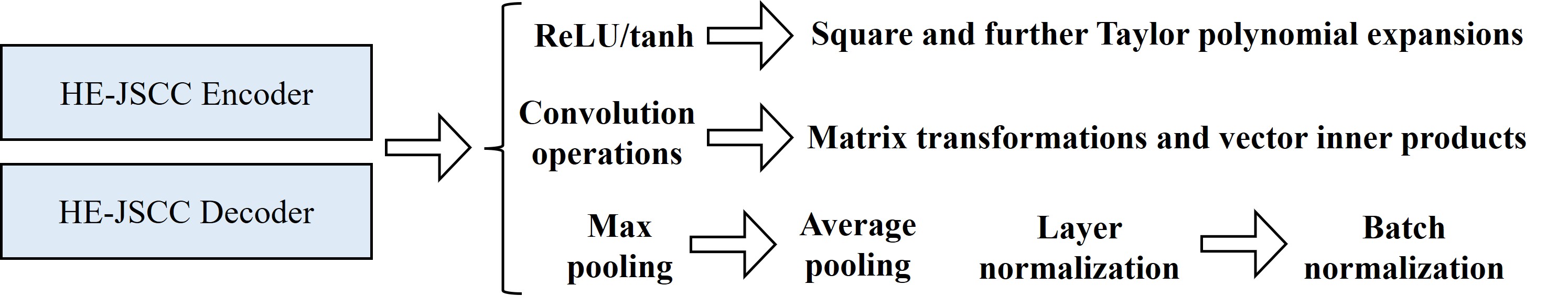}\label{31}}
\hfill

\subfloat[Offline training of the proposed HE-JSCC scheme.]{\includegraphics[width=150mm]{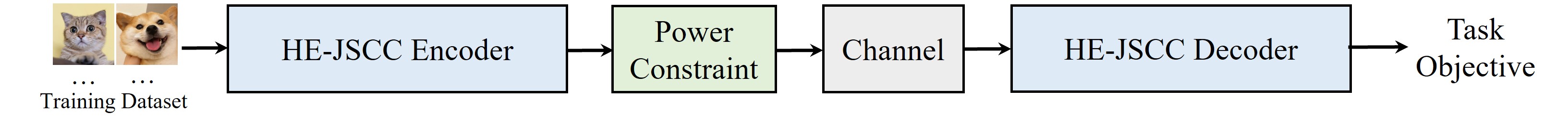}\label{32}}
\hfill

\subfloat[Online inference of the proposed HE-JSCC scheme.]{\includegraphics[width=180mm]{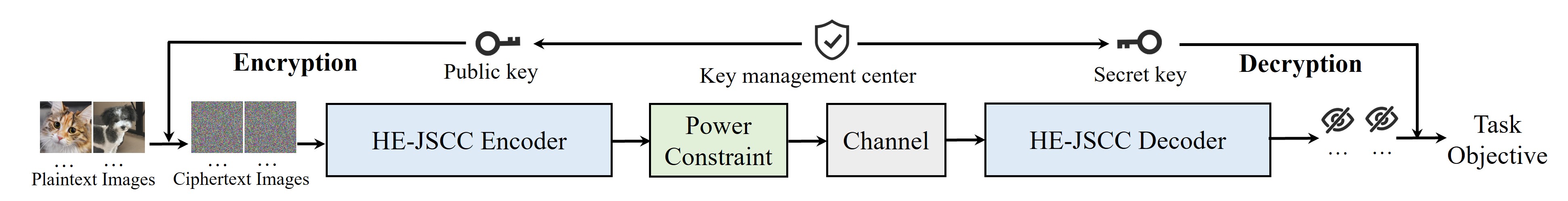}\label{33}}

\caption{The proposed HE-JSCC scheme, where (a) illustrates the modification of the HE-JSCC model, (b) illustrates the offline training stage, and (c) illustrates the online inference stage.}
\label{fig3}
\end{figure*}

\subsection{The Designed HE-JSCC Model}
As mentioned before, we need to modify the structure of JSCC models to support homomorphic operations. As illustrated in Figure \ref{31}, the designed HE-JSCC model is realized as follows:
\begin{itemize}
\item \textbf{Activation Function:}
In traditional JSCC models, activation functions commonly employ nonlinear operations. However, certain nonlinear functions pose compatibility challenges with HE. For instance, the Rectified Linear Unit (ReLU) activation function, which relies on comparison operations, and the hyperbolic tangent (tanh) function, involving exponential operations, are inherently incompatible with homomorphic computations. To address this limitation, motivated by \cite{falcetta2022privacy}, we substitute these conventional nonlinearities with the squared activation function $f(x) = x^2$, as its operations are limited to addition and multiplication and are natively supported by HE frameworks. While this substitution introduces an approximation, its accuracy can be systematically improved through Taylor polynomial expansions, enabling a balance between computational feasibility and model performance.
\item \textbf{Convolutional Layer:}
Within traditional convolutional layers, a kernel slides systematically over the input tensor. In HE-JSCC model, these convolution operations can be recast using linear algebra constructs: matrix transformations and vector inner products that align with HE requirements. However, it introduces substantial computational complexity, as each scalar in transformed matrices becomes an encrypted entity. To counteract this inefficiency, inspired by \cite{lou2019she}, we employ shift operation to reduce the number of operations needed.
\item \textbf{Pooling Layer:}
Pooling layers in traditional JSCC models are typically employed to reduce computational complexity while preserving critical semantic features. However, the max pooling operation relies on comparison operators that are incompatible with HE schemes. To mitigate this limitation, alternative pooling strategies can be adopted. Specifically, we substitute max pooling with average pooling, as this approach eliminates the need for comparative operations and requires only multiplication of ciphertexts by pre-determined constants, which are fully supported by HE frameworks.
\item \textbf{Layer Normalization:}
Traditional layer normalization techniques compute the mean and standard deviation of input data dynamically during inference to scale and shift feature values. However, in FHE frameworks, direct computation of these statistical measures on encrypted data is infeasible. Motivated by \cite{rovida2024encrypted}, we employ batch normalization to circumvent this limitation by utilizing precomputed statistics derived from the training dataset. These statistics are fixed during model training and integrated as constant parameters in the inference stage. This approach eliminates the need for statistical computation on ciphertexts, ensuring compatibility with FHE constraints while preserving normalization benefits.
\end{itemize}

\subsection{Steps of The Proposed HE-JSCC Scheme}

We outline a generalized training and inference procedure of the proposed HE-JSCC scheme as follows:

\textit{Offline Training:} As illustrated in Figure \ref{32}, the offline training includes the following steps:
\begin{enumerate}
\item \textit{Step 1:} Prepare image datasets tailored for training objectives.
\item \textit{Step 2:} Define SNR values, model hyperparameters (e.g., learning rate, batch size), and task-specific objective functions (e.g., cross-entropy loss for classification tasks, mean squared error for generative tasks). Optimize HE-JSCC parameters using backpropagation and gradient descent, iteratively minimize the specified loss function to refining the HE-JSCC model.
\item \textit{Step 3:} Adjust SNR values and re-optimize the corresponding HE-JSCC parameters to enhance robustness across varying channel conditions.
\end{enumerate}

\textit{Online Inference:} As illustrated in Figure \ref{33}, upon determining the HE-JSCC parameters, they are deployed across both the transmitter and receiver to enable online inference. It is emphasized that while the security of parameter sharing during deployment is not the privacy concern of this paper, it can be addressed through secure channel transmission. Key distribution and updates are managed by a centralized key management center, which handles both public and private keys. The transmitter utilizes the public key to encrypt source data and employs the HE-JSCC encoder for semantic feature extraction and compression. Conversely, the receiver applies the HE-JSCC decoder for semantic reconstruction, complemented by private-key decryption to recover plaintexts.

The proposed HE-JSCC scheme has the following advantages:

\begin{itemize}
\item \textbf{Efficient Training:} During the training process of the HE-JSCC model, samples can be unencrypted plaintext images. Since HE introduces additional computational complexity, constructing a plaintext image training set significantly reduces time consumption and enhances training efficiency. Furthermore, plaintext training enables visualization of intermediate features, allowing verification of whether the semantic extraction process accurately focuses on critical semantic regions.
\item \textbf{Flexible Support for Key Management:} During the online inference process, the SemCom system can dynamically adjust the key update strategy according to actual requirements. Benefiting from the homomorphic computation properties supported by the proposed HE-JSCC model, the dynamic adjustment of keys does not negatively impact the performance of SemCom. When transmitting highly sensitive data, the system can effectively resist potential eavesdropping attacks by increasing the key update frequency. When transmitting data with low privacy requirements, the key update frequency can be reduced to lower the overhead of key distribution. When transmitting non-protected data, the encryption process can be completely skipped, allowing raw image data to be directly input into the HE-JSCC model for E2E transmission.
\end{itemize}

\subsection{Simulation Results}

\subsubsection{Classification Task}

\paragraph{Dataset}
We employ MNIST dataset\footnote{\url{https://yann.lecun.com/exdb/mnist/}} and Fashion-MNIST dataset\footnote{\url{https://github.com/zalandoresearch/fashion-mnist}} for performance verification. Both of the datasets consist of 70,000 28 $\times$ 28 gray-scale images, with 60,000 images designed for training and 10,000 for inference, These images represent a 10-class classification problem, specifically, the 10 digits in MNIST and 10 fashion products in Fashion-MNIST.

\paragraph{HE-JSCC Model Parameters}
The HE-JSCC encoder includes the convolutional layer with the output size of 1$\times$28$\times$28, the square layer with the output size of 1$\times$28$\times$28, the pooling layer with the output size of 1$\times$14$\times$14, and the flatten layer with the output size of 1$\times$196. The Rician channel is modeled as fully connected layers. The HE-JSCC decoder is composed of fully connected layers with the output size of 10.

\begin{table*}
    \renewcommand{\arraystretch}{1.2}
    \centering
    \begin{tabular}{|c|c|c|c|c|c|c|c|c|c|}
    \hline 
    \multicolumn{2}{|c|}{\textbf{Datasets}} & \multicolumn{4}{c|}{\textbf{MNIST}} & \multicolumn{4}{c|}{\textbf{Fashion-MNIST}} \\
    \hline
    \multicolumn{2}{|c|}{\multirow{2}{*}{\textbf{Scheme}}} & \multirow{2}{*}{\textbf{Scheme 1}} & \multirow{2}{*}{\textbf{Scheme 2}} & \multicolumn{2}{c|}{\textbf{Proposed HE-JSCC Scheme}} & \multirow{2}{*}{\textbf{Scheme 1}} & \multirow{2}{*}{\textbf{Scheme 2}} & \multicolumn{2}{c|}{\textbf{Proposed HE-JSCC Scheme}}  \\
    \cline{5-6} 
    \cline{9-10}
    \multicolumn{2}{|c|}{} & & & \textbf{Training} & \textbf{Inference} & & & \textbf{Training} & \textbf{Inference} \\
    \hline
    \multirow{6}{*}{\textbf{\makecell{Classification\\Accuracy \\(Different SNRs)}}} & \textbf{0 dB} & 10.09\% & 88.73\% & 81.35\% & 81.37\% & 10.00\% & 80.50\% & 74.39\% & 74.39\% \\
    \cline{2-10} 
     & \textbf{5 dB} & 10.14\% & 90.32\% & 87.61\% & 87.60\% & 10.03\% & 81.92\% & 79.49\% & 79.47\% \\
    \cline{2-10} 
     & \textbf{10 dB} & 69.94\% & 91.13\% & 89.88\% & 89.60\% & 64.52\% & 82.53\% & 80.18\% & 80.18\% \\
    \cline{2-10} 
     & \textbf{15 dB} & 70.08\% & 91.80\% & 90.98\% & 90.96\% & 64.54\% & 83.14\% & 81.70\% & 81.70\% \\
    \cline{2-10} 
     & \textbf{20 dB} & 82.46\% & 92.11\% & 91.80\% & 91.82\% & 70.08\% & 83.23\% & 82.08\% & 82.08\% \\
    \cline{2-10} 
     & \textbf{25 dB} & 91.29\% & 92.35\% & 92.31\% & 92.33\% & 83.08\% & 83.53\% & 82.42\% & 82.41\% \\
    \hline
    \multicolumn{2}{|c|}{\textbf{Compression Ratio}} & 25.00\% & 25.00\% & 25.00\% & 25.00\% & 25.00\% & 25.00\% & 25.00\% &  25.00\%\\
    \hline
    \multicolumn{2}{|c|}{\textbf{Comp. Time}} & 0.4197 s & 0.000173 s & 0.000172 s & 73.84 s & 0.4236 s & 0.003 s & 0.000266 s &  59.23 s\\
    \hline
    \end{tabular}
    \caption{Simulation results for the classification task. \textbf{Scheme 1:} The source-channel separation coding model with AES encryption, where the source encoding and channel encoding are realized respectively by JPEG 2000 and LDPC, and the classification is achieved by the ResNet delopyed at the receiver. \textbf{Scheme 2:} The traditional JSCC model without modification, where the datasets are plaintext images. \textbf{The Proposed HE-JSCC Scheme:} Plaintext and ciphertext images are used during the training and inference stages of the proposed HE-JSCC scheme, respectively.}
    \label{tab2}
\end{table*}

\paragraph{Results and Analysis}
The simulation results are provided in Table \ref{tab2}. 
Firstly, due to the ``avalanche effect'', the results of scheme 1 has the worst classification accuracy, especially in low SNRs.
Secondly, the almost consistent results of training and inference stages of the proposed HE-JSCC scheme demonstrates its effectiveness in extracting semantic information from ciphertext images.
Thirdly, the comparison of the results between scheme 2 and the proposed scheme shows that, to support homomorphic operations, there are no nonlinear operations in the HE-JSCC model, which results in a loss of classification accuracy. However, as SNR increases, this loss gradually becomes negligible. For instance, when SNR = 0, there is a 7.38\% difference for MNIST datasets and a 6.11\% difference for Fashion-MNIST datasets in classification accuracy, whereas at an SNR of 25, the difference in classification accuracy is only 0.04\% for MNIST datasets and 1.11\% for Fashion-MNIST datasets. 
Fourthly, as expected, HE introduces significant overhead in terms of computation time.

\subsubsection{Generation Task}
\paragraph{Dataset}
We employ CIFAR10 dataset\footnote{\url{https://www.cs.toronto.edu/~kriz/cifar.html}} for performance verification. It consists of 60,000 32 $\times$ 32 gray-scale images, with 50,000 images designed for training and 10,000 for inference.


\begin{figure}
\centering
\includegraphics[width=0.5\textwidth]{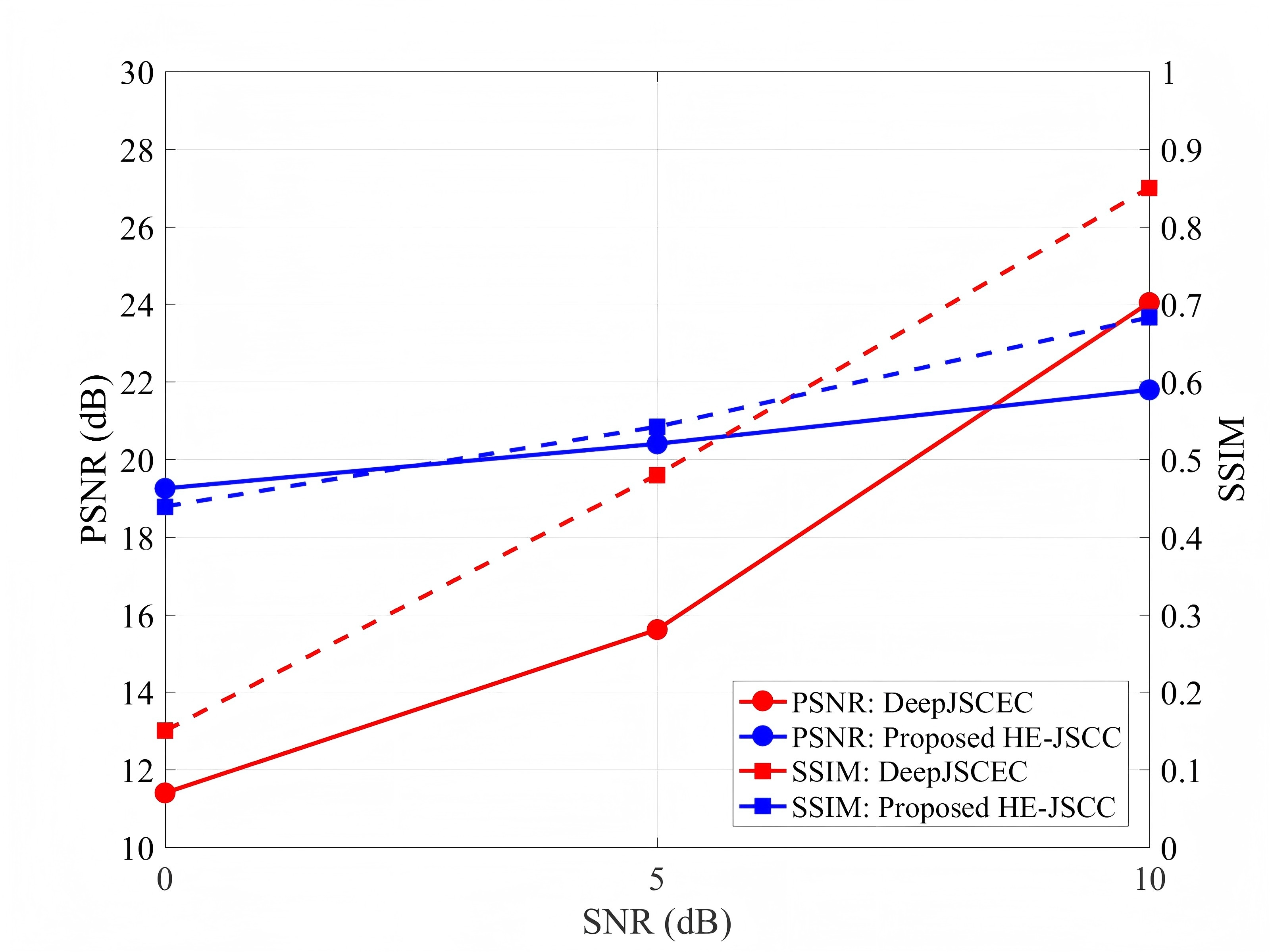}
\caption{The performance of PSNR and SSIM versus different SNRs, where DeepJSCEC \cite{tung2023deep} is employed as the comparison scheme.}
\label{fig4}
\end{figure}

\paragraph{Results and Analysis}
Figure \ref{fig4} compares the performance with DeepJSCEC \cite{tung2023deep} in Peak Signal-to-Noise Ratio (PSNR) and Structural Similarity Index (SSIM). The proposed HE-JSCC model jointly learns the encrypted semantic information and channel robustness.
Thus, in low SNR environments, such as SNR = 0 dB, the proposed HE-JSCC scheme has much better generation performance than DeepJSCEC. 
To support homomorphic operations, there are no nonlinear operations in the HE-JSCC model, which results in a loss of PSNR and SSIM. Therefore, in high SNR environments, such as SNR = 10 dB, DeepJSCEC has better performance.


\section{Future Research Direction}
\subsection{Improvement of Encryption Efficiency}

From a complexity viewpoint, the runtime bottleneck of the proposed HE-JSCC model is the nonlinear component, the polynomial activation $f(x)=x^2$. Linear modules such as convolutions and fully connected layers with fixed weights, average pooling, and fixed BatchNorm are affine and comparatively lightweight, so we omit their complexity. If $M$ encrypted elements pass through the square activation, the activation stage costs $T_{\text{act}} \approx M \cdot O(kn\log n)$, which governs the overall inference time.
To address this issue, we plan to adopt semantic importance guided partial encryption. A lightweight saliency or segmentation detector ranks pixels or feature blocks by task relevance, and only a region of interest with fraction $\rho$ is encrypted. The background remains unencrypted and is obfuscated in plaintext using coarse quantization or downscaling to limit leakage.

\subsection{Enhancement of HE-JSCC Models}
\subsubsection{Optimization of Model Parameters}
Including the Learned Perceptual Image Patch Similarity (LPIPS) term in the objective function guides optimization toward perceptual similarity in a deep feature space, which better preserves textures, edges, and object shapes under low bit rate and deep HE settings and reduces oversmoothing in reconstructions.  Therefore, future optimization can target semantic fidelity and task performance under constrained bandwidth, HE computation depth, and latency budgets. A composite loss could be adopted as $\mathcal{L}_{\text{dist}}=\lambda_{1}\mathrm{MSE}(x,\hat{x})+\lambda_{2}\bigl(1-\mathrm{SSIM}(x,\hat{x})\bigr)+\lambda_{3}\mathrm{LPIPS}(x,\hat{x})$, where the variables include model parameters $\theta$, compression ratio $\rho$, and effective bit rate $R$. A constrained multiobjective formulation will be employed as $\min_{\theta,\rho,R}\ \mathbb{E}_{\text{data,SNR}}[\mathcal{L}]$ that is subject to $R\le B$, $\rho\in\mathcal{R}(B)$, $|\theta|\le P_{\max}$, $T\le T_{\max}$, and $D\le D_{\max}$, where $B$ is the bandwidth budget, $P_{\max}$ is the upper limit on the number of parameters, $T_{\max}$ is the upper limit on latency, and $D_{\max}$ is the upper limit on HE computation depth.


\subsubsection{Expansion to Other JSCC Schemes}
Recently, researchers have designed various JSCC schemes, such as NTSCC and SP-EDNSC, which can deliver stronger performance in the cleartext domain. However, integrating these schemes with HE presents notable challenges. Non-polynomial activations and normalization layers are not directly supported by HE and call for high-degree polynomial surrogates. Additionally, quantization, entropy modeling, and discrete selections such as argmax or top-k rely on comparisons and rounding that are unfriendly to HE operations. These substitutions raise multiplicative depth, amplify noise growth, increase latency, and can distort semantic distributions, which complicates clean verification of the role of HE in SemCom. The combination of HE and these JSCC schemes is an important direction for future work. Promising paths include low-degree replacements for non-polynomial components, differentiable proxies for quantization, and entropy models that preserve semantic structure.

\subsection{Combination with Other Privacy Protection Techniques}
If the HE-JSCC model lacks sufficient generalization capability, it will still fail to effectively protect the privacy of transmitted data. Therefore, in the future, it is necessary to integrate other techniques to enhance the HE-JSCC model's generalization ability. For instance, adversarial training generates adversarial examples by adding small and imperceptible perturbations to the input data, allowing the model to encounter a wider variety of samples during training and thus improving its generalization. Adversarial training can help the HE-JSCC model find a flatter loss surface, where the loss function varies more smoothly in the parameter space. A flatter loss surface typically indicates better generalization capabilities.

\section{Conclusions}
As a novel communication paradigm, SemCom significantly reduces bandwidth requirements and enhances the reliability of communication systems through efficient and accurate semantic extraction, transmission, and recovery. However, the urgent security issues in SemCom, such as privacy leakage of semantic data and damage to semantic data integrity, need to be addressed. In this paper, we have first revisited existing encryption schemes for traditional mobile systems. Then, we have analyzed the limitations of applying the state-of-the-art schemes to secure SemCom. Due to the homomorphic operation characteristic of HE, we have applied SIFT to verify the extractable semantic features can be preserved in homomorphic encrypted ciphertext. Based on this finding, we have further analyzed what SemCom can benefit from HE. Moreover, we have proposed the HE-JSCC scheme and presented the simulation results for image classification and image generation tasks. We have further provided future research directions for researchers in related fields.

\bibliography{ref.bib}
\bibliographystyle{IEEEtran}

\vfill

\end{document}